# Influence of Bubbles on the Energy Conversion Efficiency of Electrochemical Reactors

Andrea Angulo[1], Peter van der Linde[2], Han Gardeniers[2], Miguel Modestino[1,*], David Fernández Rivas[2,*]

[1]Department of Chemical and Biomolecular Engineering, Tandon School of Engineering, New York University. 6 Metrotech Center, Brooklyn, NY 11201, U.S.A.

[2]Mesoscale Chemical Systems Group, MESA+ Institute and Faculty of Science and Technology, University of Twente, P.O. Box 217, 7500 AE Enschede, The Netherlands.

*Corresponding authors: modestino@nyu.edu, d.fernandezrivas@utwente.nl

ORCID 0000-0002-7504-5946 – Andrea Angulo
ORCID 0000-0002-2696-1040 – Peter van der Linde
ORCID 0000-0003-0581-2668 – Han Gardeniers
ORCID 0000-0003-2100-7335 – Miguel Modestino
ORCID 0000-0003-4329-3248 – David Fernandez Rivas

**SUMMARY**

**Bubbles are known to influence energy and mass transfer in gas evolving electrodes. However, we lack a detailed understanding on the intricate dependencies between bubble evolution processes and electrochemical phenomena. This review discusses our current knowledge on the effects of bubbles on electrochemical systems with the aim to identify opportunities and motivate future research in this area. We first provide a base background on the physics of bubble evolution as it relates to electrochemical processes. Then we outline how bubbles affect energy efficiency of electrode processes, detailing the bubble-induced impacts on activation, ohmic and concentration overpotentials. Lastly, we describe different strategies to mitigate losses and how to exploit bubbles to enhance electrochemical reactions.**

**Context & Scale**

**Electrochemical reactors will play a key role in the electrification of the chemical industry and can enable the integration of renewable electricity sources with chemical manufacturing. Most large-scale industrial electrochemical processes, including chloro-alkali and aluminum production, involve gas evolving electrodes. The evolution of bubbles at the surface of redox reaction sites often lead to the reduction of the active electrode area, the increase of ohmic resistance in the electrolyte and the formation of undesirable concentration gradients. All of these effects result in energy losses which reduce the efficiency of electrochemical systems. This review synthesizes our current understanding on the relationship between bubble evolution and energy losses in electrochemical reactors. By presenting a thorough account on the state of the research in this area, we aim to provide a common ground for the research community to improve our understanding on the complex processes involved in multiphase electrochemical systems. Increasing our knowledge on the relationship between bubbles and electrochemistry will lead to new strategies to mitigate and exploit bubble-induced phenomena leading to design guidelines for high-performing electrochemical reactors.**

# 1. INTRODUCTION

Gas evolving electrodes are encountered in a large number of electrochemical systems where electron transfer events result in the formation of products with limited solubility in the electrolyte, leading to the formation of bubbles. Three of the largest industrial electrosynthesis processes (*i.e.*, chloro-alkali, sodium chlorate and aluminum production) evolve gases at the surface of at least one of their electrodes.[1, 2] Bubbles are also formed in a multitude of emerging electrochemical processes for the sustainable production of fuels and chemicals (*e.g.,* water splitting, $CO_2$ electroreduction, or organic electrosynthesis).[3, 4, 5, 6, 7, 8] The most common gas evolving electrochemical reactions include hydrogen or oxygen evolution from water splitting, chlorine generation from chloride oxidation, direct methanol fuel cells or carbon dioxide ($CO_2$) generation from sacrificial carbon electrodes.[3, 4, 5, 9, 10] It is often the case that desired reductions or oxidations are coupled with one of these gas-evolving redox transformations as a sacrificial reaction. For example, the chloro-alkali process evolves $H_2$ in the cathode as a sacrificial reduction product, or the aluminum reduction process is coupled with the anodic oxidation of carbon to $CO_2$.

Bubbles that evolve at an electrode may result in undesired blockage of the electrocatalyst surface,[11, 12] and ion conducting pathways in the electrolyte, resulting in energy losses.[13] Although bubble nucleation and, in consequence, their associated potential variation are dependent on the operating current density, cell design (electrode geometry, interelectrode spacing) and electrolyte formulation, several studies have estimated these losses under specific conditions. In an aluminum production industrial cell, potential drops in the range of 0.09-0.35 V were reported for 10 kA cell with a bubble layer of ~ 2.1 cm,[14] while laboratory scale studies reported potential drops of 0.15-0.35 V for current densities of 300-900 mA/cm$^2$.[15, 16] In chloro-alkali processes at industrially relevant current densities (>600 mA/cm$^2$), bubbles attached to the electrodes accounted for potential drops in the range of 0.6 to 0.9 V, ~20% of the total cell voltage.[17, 18] In water electrolysis, which will be the focus of this review, an increase of 15-18% of current density, corresponding to a potential decrease in the range of 10-25% has been reported.[19] In all-vanadium redox flow batteries oxygen and hydrogen evolution cells have found to negatively impact the cell performance due to partial obstruction of the electrolyte flow, reduction of the electrocatalytic area and reduction of the mass and charge transfer coefficients.[20, 21] The somewhat similar problem, formation of droplets in systems where the reactants are in the gas phase is also found. In particular, in proton exchange membrane fuel cells (PEMFC) the formation of water droplets is a concern as they can cause flooding of the membrane preventing, access of the reactant gases towards the electrodes.[22]

Additionally, bubbles in electrochemical processes are known to induce convection and enhance mass transfer rates.[23, 24, 25, 26] Improving the efficiency of existing or emerging electrochemical processes, will require a detailed understanding of the behavior of bubbles in electrochemical systems. This understanding can lead to the development of methods to mitigate the detrimental effects of bubbles on energy conversion efficiency and stability, at the same time exploiting the benefits of bubble-induced phenomena.

This review introduces the reader to a growing body of interdisciplinary literature on the effects of bubbles in the energy and mass transfer efficiency of electrochemical systems. While previous reviews have focused on the mathematical modeling of electrochemical gas evolving systems or discussed the interactions of bubbles and electrocatalyst surfaces,[19, 27, 28] we provide a complementary viewpoint covering electrochemical engineering aspects of devices for the production of electrofuels and chemicals. Specifically, we will focus on hydrogen evolution in water electrolyzers given its relevance to energy research but most of the equations shown in the following sections are applicable for most gas evolving system. We start by discussing the dynamic phenomena that lead to bubble formation, growth and detachment, followed by a detailed account on the energy losses caused by the presence of bubbles in electrochemical systems. We then present a set of strategies to manage the behavior of bubbles and exploit their properties in gas evolving electrodes. This thorough (non-exhaustive) account of the interactions between bubbles and electrochemical processes is aimed at accelerating the development of electrochemical systems with enhanced operational efficiencies.

## 2. BUBBLE EVOLUTION DYNAMICS

Electrochemical bubble evolution is defined as the nucleation and growth at, and detachment from electrodes (see Figure 1).[29] These phenomena have been studied in batch,[30] and flow[3] reactors with the goal of understanding the relationship between the structure multiphase electrolytes and the performance of electrochemical reactions

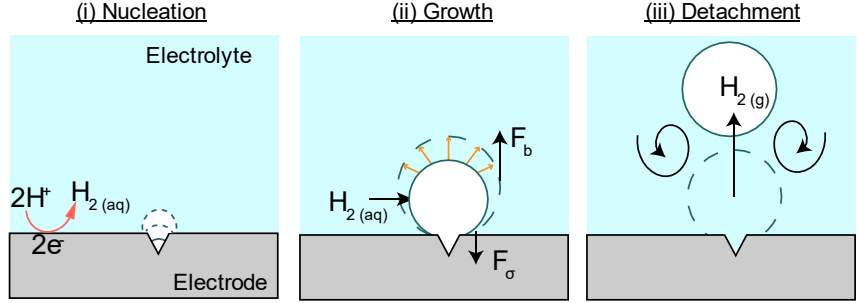

**Figure 1. The various stages of bubble evolution**
(i) The nucleation, (ii) growth, and (iii) detachment of bubbles on a gas evolving electrode surface. Nucleation occurs typically on cracks and crevices in the electrode surface, after which the bubble grows by taking in gas from the dissolved gas boundary layer. Bubbles detach when the buoyancy force overcomes the interfacial tension force. The departing bubble induces convection in the liquid indicated by the spirals.

**2.1 Nucleation**

Bubble nucleation during electrolysis is driven by an increased chemical potential of dissolved gas molecules near the electrode surface (see Figure 2). The equilibrium or saturation concentration of dissolved gas in a liquid, $C_{sat}$, is proportional to the partial pressure $P$ of the gas acting on a liquid surface (Henry's law):

$$C_{sat} = P K_H(T) \qquad \text{(Eq. 1)}$$

Henry's solubility constant $K_H$ is a decreasing function of temperature $T$ specific to each liquid-gas pair.[31, 32, 33, 34] In electrochemical processes, nucleation takes place when the concentration of dissolved gas near the electrode surface $C_0$, becomes large enough (see Figure 2). For simplicity, in what follows, we will consider the case when only a single gas species is present. The level of gas supersaturation of a liquid at a pressure $P_0$ can then be expressed as:

$$\zeta = \frac{C_o - C_{sat}}{C_{sat}} = \frac{C_o}{K_H P_0} - 1 \qquad \text{(Eq. 2)}$$

The liquid is said to be supersaturated if $\zeta > 0$ ($C_0 > C_{sat}$) and undersaturated if $\zeta < 0$ ($C_0 < C_{sat}$).[35] The concentration of dissolved gas in the liquid immediately adjacent to the interface of a spherical bubble, $C_b$, can be obtained by combining Eq. 1 with the Laplace-Young equation:

$$C_b = K_H P_i = K_H \left(\frac{2\gamma}{r} + P_0\right) \qquad \text{(Eq. 3)}$$

Here $P_i$ denotes the gas pressure inside the bubble with radius $r$, and $\gamma$ is the surface tension (the energy required to increase the surface of a liquid due to intermolecular interactions)[36] of the gas-liquid interface. The quantity $\frac{2\gamma}{r}$ is known as the Laplace pressure valid for small bubbles, assuming there is a constant curvature. This is typically the case in microfluidics where Bond numbers are small enough, meaning that the surface tension dominates over buoyancy.

A bubble grows when $C_0 > C_b$; the condition that drives mass transfer of dissolved gas from the liquid to the bubble. Supersaturation is therefore a necessary condition for bubble nucleation and subsequent growth, except in the case where multiple gas species are present.[30] The effect of multiple gases on electrolysis bubbles has not been studied in detail and is out of the scope of this review.

In practice, the supersaturation value allows quantifying the tendency of a system to produce bubbles. The gas supersaturation near an electrode surface depends on several factors, such as reactor and electrode geometry and wettability, the current density, the flow field, the time elapsed since the start of electrolysis, the amount of bubble coverage, the electrolyte properties, etc. Typical supersaturation values for hydrogen electrolysis at which bubble nucleation are experimentally observed are $\zeta \sim 5 - 10$,[37] these are to be contrasted with theoretical considerations of homogeneous bubble nucleation of $\zeta \sim 10^5$,[38] or calculations for hydrogen evolution in pure water at 25 °C which range between 8-24.[39] The saturation concentration of Hydrogen at 28 °C is ~0.75x10$^{-3}$ mol/L[40] with a $K_H$~7.7x10$^{-6}$ mol/m$^3$ Pa.[41] In the literature there are other ways to report supersaturation values, e.g. in units of pressure[42, 43, 44] or as the ratio between $C_0$ and $C_{sat}$.[40, 45]

While homogeneous nucleation in the electrolyte is possible, in practice, bubbles tend to form at heterogeneous interfaces (*e.g.*, crevices in the electrode surface or impurities in the liquid phase) where the thermodynamic barrier for nucleation is lower.[38, 46, 47, 48, 49, 50] Even with relatively simple liquids such as pure water, the experimental conditions for homogeneous nucleation have been difficult to match theoretically due to the presence of impurities or crevices.[51]

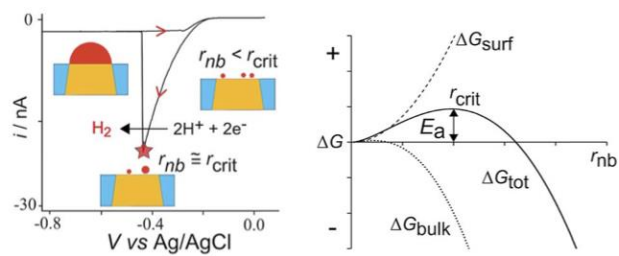

**Figure 2. (Left) Cyclic voltammogram for H$^+$ reduction in 0.5 M H$_2$SO$_4$ at a 33 nm radius Pt nanodisk electrode (500 mV s$^{-1}$).** The peak (red star) corresponds to a phase change as a nucleus grows into a stable nanobubble covering the electrode. The schematics indicate spontaneous formation of nuclei near the electrode's surface. For high enough $H_2$ electrogeneration a supersaturation at the electrode/electrolyte interface is reached that allows the existence of a nucleus of critical size, which subsequently covers the electrode as it grows **(Right) The total free energy of formation of a gas bubble nucleus is plotted against the radius of a bubble r$_{nb}$.** The maximum in *ΔG* corresponds to the critical radius, *r$_{crit}$*, with an activation energy barrier of *E$_a$*. The growth of a nucleus into a stable bubble with radius *r$_{nb}$*, is energetically favorable for bubble radii greater than *r$_{crit}$*. Modified with permission from German, S.R., et al., *Electrochemistry of single nanobubbles. Estimating the critical size of bubble-forming nuclei for gas-evolving electrode reactions.* Faraday Discuss., 2016. **193**: p. 223-240. Copyright 2016, Royal Society of Chemistry.[52]

The formation of bubbles is explained by classical nucleation theory (CNT), that defines the activation energy for bubble formation in terms of the cohesive force of the liquid, assuming that the gas nucleus volume is large enough so that bulk thermodynamic properties apply.[53] This implies that nucleation is defined as the event in which a critical nucleus size of the gas bubble, *r$_{crit}$*, is formed.[52] The sum of the energy required to form a new interface and the energy gained as dissolved gas is transferred to the newly created phase, is defined as the free energy of formation of a gas nucleus in solution, Δ*G$_{tot}$*. Figure 2(*right*) shows Δ*G$_{tot}$* as a function of the bubble's radius, *r$_{nb}$*. The gas-liquid interface free energy is $\Delta G_{sur} = 4\pi r_{nb}^2 \gamma$, where *γ* is the surface energy of the gas-liquid interface (*i.e.*, surface tension). The bulk component term is proportional to the bubble volume $4/3\pi r_{nb}^3$ and to the free energy difference between the gaseous and dissolved state in that volume, Δ*G$_V$*, leading to:

$$\Delta G_{tot} = 4\pi r_{nb}^2 \gamma + \frac{4}{3}\pi r_{nb}^3 \Delta G_V \qquad \text{(Eq. 4)}$$

As observed in Figure 2(*right*), the maximum Δ*G$_{tot}$* is defined as the activation barrier *E$_a$* for bubble nucleation. It can be calculated as the surface energy required to form a spherical cap bubble at a critical radius *r$_c$*:[54]

$$E_a = \frac{4}{3}\pi r_c^2 \gamma \phi(\theta) \qquad \text{(Eq. 5)}$$

where $\phi(\theta) = (1 + cos\theta)^2(2 - cos\theta)/4$, *θ* being the contact angle (through the liquid phase) between the gas-liquid interface and the solid substrate.[54]

We may define a critical radius for diffusive stability of a bubble in a supersaturated solution (not growing or shrinking), below which the bubble dissolves due to surface tension effects (i.e., the Laplace pressure, cf. Equation 3). This critical radius occurs when the interfacial concentration equals the bulk concentration in the surrounding liquid: $C_b = C_0$. Equating then Equation 2 and 3, we get:[35]

$$r_c = \frac{2\gamma}{P_0 \zeta} \qquad (Eq.\ 6)$$

A relevant parameter for heterogeneous bubble formation phenomena is the wettability of a surface that can be described by the spreading coefficient, defined as:[36]

$$S = \gamma_{SG} - \gamma_{SL} - \gamma. \qquad (Eq.7)$$

where $\gamma_{SG}$ is the surface energy of the solid/gas interface, $\gamma_{SL}$ the solid/liquid interface, and $\gamma$ is the liquid/air interface.[36] The work $\delta W$ necessary to displace the contact line along a direction parallel to the plane $\delta x$ is:

$$\delta W = (\gamma_{SG} - \gamma_{SL})\delta x - \gamma \cos\theta\, \delta x;\ or\ \cos\theta = \frac{S}{\gamma} + 1. \qquad (Eq.\ 8)$$

Young's equation is found by considering that the work at equilibrium must be zero and written as:

$$\gamma_{SG} - \gamma_{SL} = \gamma \cos\theta \qquad (Eq.\ 9)$$

where $\theta$ is the contact angle through the liquid phase. For positive S, total wetting, is said when the system forms a film that covers the solid. For $S \leq -2\gamma$, non-wetting, the liquid minimizes the contact with the solid. For Intermediate cases, partial wetting, the contact line is determined by Young's relation.[36]

The nucleation of nanometer-scale bubbles was thought not to be possible due to large Laplace pressures. However, it has been demonstrated that pinning of the three-phase contact line at surface heterogeneities provides stability for the existence of nanobubbles.[55] As a limitation, the nucleation rates obtained from CNT are known to differ substantially from experimental measurements.[56, 57] In electrolysis studies, often the theoretical nucleation rates and supersaturation levels for nucleation are higher than those observed experimentally.[38, 58, 59, 60] This discrepancy has been attributed to surface tension effects not accounted for in theoretical predictions,[59] or to the presence of non-spherical bubble shapes which would affect the supersaturation threshold for nucleation.[61]

While heterogeneous bubble nucleation is favored thermodynamically, a few reports have suggested the occurrence of homogeneous nucleation. Hydrogen bubbles were observed to form near a mercury pool electrode surface.[37] The homogeneous nucleation was attributed to an intense electric field (-2.3×10$^8$ V m$^{-1}$) in the electric double layer, originating from the polarization of the liquid molecules lowering the local surface tension, which reduced the nucleation energy barrier. Similarly, nucleation was observed near the proximity of a gold electrode (25 nm diameter) with *in situ* transmission electron microscopy, suggesting an homogeneous nucleation event attributed to the electrode wettability (Figure 3).[50] The challenges to image homogeneous nucleation events with the required spatial-temporal resolution are significant and have prevented the scientific community to gain insights into this phenomenon. Achieving homogeneous nucleation is challenging due to the presence of small impurities or surface heterogeneities that tend to serve as nucleation sites. A separation of 7 nm in Figure 3 could indicate that the frame rate available to the researchers was not sufficient to capture the onset of bubble evolution and hence it is difficult to attest without doubt that the observe bubble was generated through a homogeneous nucleation process.

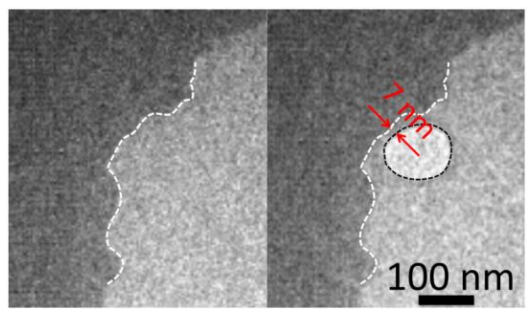

**Figure 3. In situ transmission electron microscopy images of the nucleation of a hydrogen bubble next to a gold electrode, outlined by the dashed line**
The left image shows the electrode (in dark gray, a white dashed line marks its outer edge) before nucleation, the right image shows the bubble (in light gray circled by a black dashed line) adjacent to the electrode. The time step between the two images is 340 ms and the distance between bubble and electrode is 7 nm. Reproduced with permission from Liu, Y. and S.J. Dillon, *In situ observation of electrolytic H2 evolution adjacent to gold cathodes. Chem. Commun.,* 2014. 50: p. 1761-1763. Copyright 2014, Royal Society of Chemistry.[50]

Whereas in practical electrochemical systems the uncontrolled formation of multiple bubbles is unavoidable, the research community has developed several techniques to isolate bubble nucleation events and study their effects on gas evolving electrodes. Studying single nucleation events can avoid measurement interference from convection upon detachment of other bubbles,[23, 62] from bubble coalescence,[63] and from bubble Ostwald ripening events.[64] This isolation can be achieved by using nano- and micro-sized electrodes.[26, 65, 66, 67] These structured electrodes can help to understand the dynamics of dissolved gas boundary layers and its effect on nucleation,[32] bubble growth,[30] and mass transfer under spatial confinement.[68, 69]

Structures such as inverted pyramidal structures have been used to exert control over the location of bubble nucleation. When such structures are created in electrodes, the dissolved gas concentration fields overlaps in the apex of the cavity, leading to oversaturated liquid and a larger chemical potential locally,[32] which increases nucleation probability within the cavity. Similarly, the onset location for bubble nucleation can be structurally defined with needle electrodes.[70] Under this geometry, the convergent electric field at the tip of the electrode controls the charge transfer rate and results in a single active site for successive bubble nucleation.

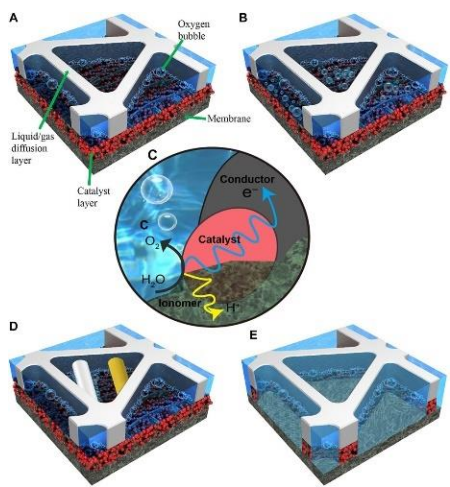

**Figure 4. Schematic of opening-scale electrochemical reactions occurring in the anode of a PEMEC.** (A) True electrochemical reaction phenomena as revealed in this study. (B) Conventional perception of electrochemical reactions. (C) TPB electrochemical reaction. (D) Introduction of a conductive (white) and a non-conductive (yellow) wire that confirms that bubbles are nucleating at the site of the electrochemical reaction. (E) Suggested future design for CLs in PEMECs: the catalyst is only deposited on the lands of LGDLs. (permission pending) from Mo, J., et al., *Discovery of true electrochemical reactions for ultrahigh catalyst mass activity in water splitting.* Science Advances, 2016. **2**.

In proton exchange membrane electrolyzer cells (PEMECs), the use of a high speed imaging and microscale visualization system allowed the identification of nucleation events at the site of the electrochemical reaction.[71] For the oxygen evolution reaction in the anode it was observed how electrochemically generated bubbles nucleated mainly along the catalyst layer (CL) - liquid/gas diffusion layer (Figure 4.A) and not across the entire exposed catalyst surface as it was previously thought (Figure 4.B).[71, 72] Based on these findings, a new design for CLs in PEMECs that maximizes catalyst utilization is proposed, in which catalyst is only deposited in the interface of the PEM and the substrate (Figure 4.E).

**2.2 Growth**

The growth or shrinkage of a bubble attached on a surface during electrolysis can give information about the chemical potential in its vicinity. If the dissolved gas concentration is higher than the saturation concentration ($\zeta > 0$), then a bubble will grow; in contrast, it will shrink if the surrounding electrolyte is undersaturated ($\zeta < 0$). There are several other aspects that influence growth rate, which can include geometrical factors (presence of walls or other types of confinement), temperature, surfactants, the presence of other bubbles, among others.[63]

Three growth regimes occur during bubble evolution in electrolysis, each dominated by a different force or process. The initial stage of growth is governed by inertia imposed by the liquid surrounding the bubble, and lasts around 10 ms.[65] This stage of growth is characterized by a growth rate that can be described by $r_b = bt$, with $r_b$ the radius of the bubble, $b$ a growth coefficient, and $t$ the time.

The second stage is governed by mass transfer of dissolved gases to the bubble and the growth rate can be described by $r_b = bt^{1/2}$. Diffusion controlled bubble growth during electrolysis occurs once the liquid surrounding the bubble saturates such that diffusion of dissolved gas occurs from the bulk electrolyte towards the bubble. This phenomenon has been extensively covered in literature, including theoretical descriptions and experimental work of bubble growth and dissolution stages.[35, 73, 74] Diffusion controlled bubble growth is also observed in boiling[75] and pressure driven supersaturation,[76] leading to qualitative comparisons between electrolytic gas production rate curves to boiling curves.[77] For example, electrolysis curves show a hysteresis effect, where the number of activated cavities at decreasing current densities exceeds its value when increasing the current; similar to the hysteresis observed in boiling processes attributed to advancing-receding contact line phenomena.[78] This analogy between electrolytic gas evolution and boiling may not be generally valid when the limiting rate processes are not equivalent.[79] To the best of our knowledge, the deactivation of nucleation sites is a process that has not been studied in electrochemical systems. However, it may be explained by the filling of crevices with electrolyte after a bubble has been formed or during detachment (see next subsection), which can result in the deactivation of an heterogeneous nucleation site.[80]

In the third stage, bubble growth is limited by the electrochemical reaction rate, and can be described by a $r_b = bt^{1/3}$ dependence where the exponent originates from the constant rate of gas addition to the bubbles.[26, 65, 79, 81, 82] Reaction-limited mass transfer can be observed for nano- and microelectrodes where diffusion is fast due to the short distance between the gas evolving electrodes and the bubble. Although most electrolysis systems follow the growth regimes described above, deviations can occur. For example, in the case of moving gas-evolving sources such as catalytic micro-motor particles the bubble growth follows a $r_b = bt^{\sim 1/4}$ dependence.[83] Deviations from reaction- or diffusion-limited growth are also observed during bubble evolution in an underdeveloped dissolved gas boundary layer.[30]

**2.3 Detachment**

Bubble detachment is the process of a bubble unpinning from a surface.[84] The maximum theoretical bubble detachment radius for a bubble on an upward facing electrode in a stagnant electrolyte is given by Fritz's formula:[85]

$$r_d^* = \left(\frac{3r_0 \sigma \sin\theta}{2\Delta\rho g}\right)^{1/3} \quad \text{(Eq. 10)}$$

The radius of the circular area where the bubble is in contact with the electrode is $r_0$, $\Delta\rho$ is the difference in density between the liquid and gas phases, and $g = 9.81$ m s$^{-2}$ the gravitational acceleration. In the case of bubble detachment, it is also useful to consider analogies between boiling and electrolysis, in particular for low current density regimes where wettability changes are not observed and the behavior is qualitatively similar.[86, 87] Direct observations of bubble contact line detachment is an experimental challenge, particularly due to fast dynamics and accessibility to visualize the small scale in which it occurs (see Figure 5). In both cases (electrolysis and boiling) it is generally accepted that as the bubble grows, a narrowing of the neck attached to the surface takes place, followed by detachment. Then, a small gas pocket is left upon bubble detachment, which guarantees a new cycle of bubble growth and detachment. Electrolytic bubbles detaching from microelectrodes with various radii were employed to study the effect of pH on the net charge induced on the bubble interface, the influence of electrostatic interactions on detachment radii of bubbles, screening effects, and the influence of surfactants (electrostatic interaction influences are discussed in section 4).[65] It was found that the bubble departure diameter in surfactant-free electrolytes depended on pH, a point of zero charge at pH 2-3, attributed to electrostatic interactions between the electrode and double layer charges on the bubbles.

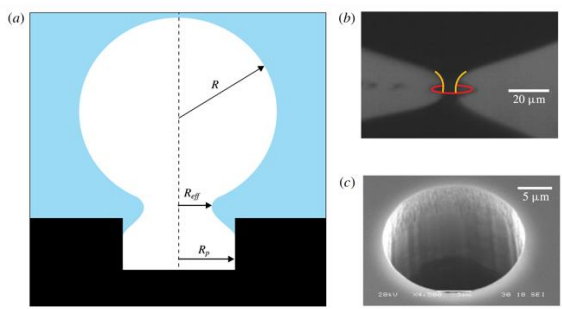

**Figure 5. A bubble attached to a hydrophobic pit. a)** The bubble detaches from a neck smaller than the pit radius; assuming perfect sphericity and neglecting the gas contained in the pit has been calculated to be smaller than 0.3%. **b)** Zoom in at the bubble neck just before detachment. The circle indicates the opening of the pit. **c)** Scanning electron microscopy image of the hydrophobic pit with black silicon coating the bottom to promote gas entrapment. Republished with permission from Soto, A.M., *Bubbles on surfaces: Diffusive growth & electrolysis*. 2019, University of Twente: Enschede. [88]

Following the assumption that bubble phenomena are similar across different experimental settings, well-controlled pressure driven bubble growth due to gas oversaturation has been compared with electrolysis.[29, 30, 76, 89] In both scenarios (*i.e.*, pressure driven and electrolysis experiments), the bubble detachment occurs at comparable radii, suggesting that the source of dissolved gasses does not have strong effects in bubble detachment.[89]

The influence of surface tension on detachment has been investigated with different gases.[85, 90, 91, 92] Dissolved gases can reduce the surface tension of the liquid phase,[93, 94] resulting in soluto-capillary flow (*i.e.*, Marangoni flow).[95] Although Marangoni flow has been thought to play an important role during electrolysis, little research has been published on the topic related to electrochemistry.[96] An experimental study employing particle tracking velocimetry allowed measuring the Marangoni flow around hydrogen bubbles generated via electrolysis[97]. It was concluded that there was a correlation between the magnitude of the Marangoni convection and the electric current, presumably due to an enhanced soluto-capillary flow originating from the larger flux of dissolved gasses, or increased Joule heating leading to thermo-capillary flow.

In-situ observations of the whole process of hydrogen evolution reactions (HERs) and oxygen evolution reactions (OERs) on micro-bubbles, simultaneously studied, led to the finding that bubble detachment diameter is inversely proportional to the flow velocity, and directly proportional to the current density.[98] Similarly, with the use of high speed microscopy imaging in a transparent PEMEC, it was shown how the detachment diameter of oxygen bubbles increased with the operating current density.[72]

Engineering the topology of the electrodes can also alter the detachment behavior of bubbles. For example, bubbles originating from an orifice on a wall detach at smaller radii with increased flow rates due to the induced shear stress.[99, 100] It is thought that the detachment of bubbles from crevices can induce liquid jet formation into the crevice, filling it, and prevent further bubble growth (leading to the hysteresis described in the nucleation subsection).[80, 101] Other methods to induce early bubble detachment are discussed in detail in section 4.

## 3. BUBBLES IMPACT ON ELECTROCHEMICAL PROCESSES

The potential drop across an electrochemical cell, $\Delta V$, is determined by the equilibrium potential of the electrochemical transformation, $E_0$, and the overpotential losses in the cell. These losses can be divided in activation overpotentials, $\eta_a$ and $\eta_c$ for the anodic and cathodic reactions, respectively, ohmic overpotentials related to the transport of ions through the electrolyte ($\eta_{ohm}$) and concentration overpotentials related to differences in electrochemical potentials across the electrolyte caused by the formation of concentration gradients ($\eta_{conc}$). [102, 103]

$$\Delta V = E_0 + \eta_a + \eta_c + \eta_{ohm} + \eta_{conc} \qquad \text{(Eq. 11)}$$

The extent of these losses is determined by both electrochemical reaction and transport (diffusion, convection, migration) processes. In the case of electrochemical transformations involving gas-evolving electrodes, the presence of bubbles can affect each of these processes and in turn influence each of the overpotential losses. To illustrate these effects, let's consider the processes involved in the hydrogen evolution reaction (HER) in acidic media: (i) transport of $H^+$ ions from the anode to the cathode by means of migration, (ii) electrochemical reduction of $H^+$ into $H_2$ at the catalytic surface of the cathode and (iii) transport of products away from the cathode by means of diffusion and forced convection.[102, 104] At sufficiently low current densities, the rate at which $H_2$ is formed is low enough to allow its diffusion into the bulk electrolyte prior to reaching the concentration threshold for bubble nucleation. Under these conditions, the overall reaction proceeds in a single phase and uniformly throughout the surface of the cathode. Overpotential losses can arise from concentration gradients in the diffusion layer due to the consumption of protons, the inherent resistivity of the electrolyte that opposes ion migration, and the reaction kinetics of the HER in the surface of the electrode (Figure 6 (a)).[102]

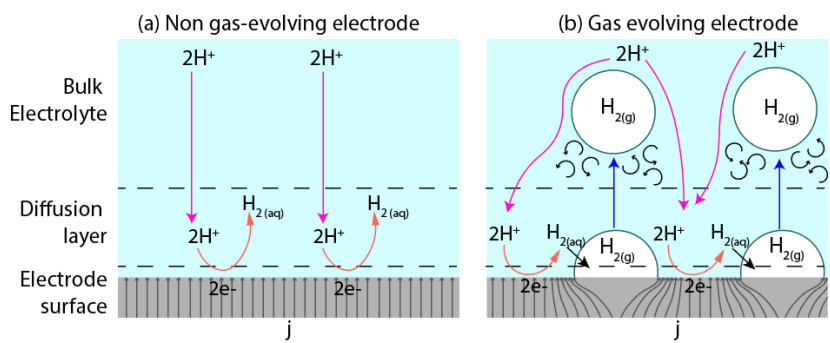

**Figure 6. Possible bubble processes in (a) single-phase and (b) gas evolving system that affect the potential losses for a hydrogen evolution electrode.** The magenta lines represent the proton migration pathways towards the cathode, the red arrows the electrochemical reaction, the blue lines show bubble detachment from the electrode's surface and the black arched arrows the local turbulence induced by detaching bubbles.

At higher current densities, the reaction occurs at a faster rate than the products can be transported away from the surface, resulting in the accumulation of dissolved gases that eventually reach a concentration above the supersaturation threshold for bubble formation (Figure 6 (b)). Under these conditions, bubbles will form on the surface of the electrodes by means of heterogeneous nucleation and grow until they reach a threshold for detachment. While the bubbles are growing, they will effectively decrease the active electrocatalytic area, inducing a non-uniform current density distribution, compressing the current lines in areas close to the bubble-catalyst-electrolyte interphase and thus leading to a local increase in current density in those areas.[105, 106] In these high current density zones, the reactants are rapidly depleted and the rate of $H_2$ generation increases, further accelerating the growth of the bubbles. Bubbles continue to grow by means of diffusion of the dissolved $H_2$ to the gas phase, and their growth induces moderate convective flows which enhance transport. Eventually, bubbles detach and move away from the electrode, inducing a wake that contributes to the local mixing of the species and decreases the concentration gradients. Lastly, bubbles block ion conduction pathways which increases the effective resistance of the electrolyte.[103, 105, 107] All these bubble-induced processes influence the efficiency of water electrolysis and have distinct effects in each of the overpotential contributions, as described in the subsections below.

**3.1 Bubble Effects on Activation Overpotential**

Also known as surface overpotential, the activation overpotential is related to the reaction kinetic losses that take place at the surface of the electrodes in each of the half-cell reactions. Typically, the Butler-Volmer equation (Eq. 12) can be used to relate the current density to the activation overpotential. In the case of HER, this relationship can be written as

$$i_c = j_0 A \left(\frac{[H^+]}{[H^+]_0}\right)^\gamma \left[\exp\left(\frac{\alpha_a F}{RT}\eta_a\right) - \exp\left(-\frac{\alpha_c F}{RT}\eta_c\right)\right] \qquad \text{(Eq. 12)}$$

where $i_c$ is the cathodic current, $j_0$ is the exchange current density, $A$ is the active area of the electrode, $\alpha_a$ and $\alpha_c$ are the anodic and cathodic apparent charge transfer coefficients, respectively, $[H^+]$ and $[H^+]_0$ are the cation concentrations at the electrode's surface and at a reference state and $\gamma$ is a reaction order that describes the strength of the contribution of the concentration term. In gas evolving electrodes, attached bubbles can effectively decrease the electrocatalytic surface, resulting in an increase in overpotential $\eta_c$ for a given current density.[108] The current density is related to the fractional bubble coverage of the electrodes, $\sigma$, by the following geometric relationship (Eq. 13):[109]

$$j = \frac{i}{A(1-\sigma)} \qquad \text{(Eq. 13)}$$

Studies on the overpotential losses due to attached bubbles as a function of exchange current density have concluded that kinetic losses dominate over ohmic losses at low current density (Figure 7), and that as $j \to 0$, the activation overpotential in this regime can be solely described by the bubble coverage fraction $\sigma$ :[105]

$$\eta_{a/c} = \frac{RT}{F} \ln \frac{1}{1-\sigma} \qquad \text{(Eq. 4)}$$

while at high current densities ohmic losses are the main source of overpotential.

Although isolating the activation from ohmic overpotential in electrochemical measurements is challenging, information on the relationship between $\sigma$ and current density may be useful to determine the active electrocatalytic area, and subsequently, predict the potential losses associated to the reduction of it due to evolving bubbles. Numerous empirical relationships between the bubble coverage and detachment diameter and current density have been derived both for stagnant,[7, 12, 109] and flowing electrolytes,[7, 12, 109, 110, 111, 112, 113] all of which are dependent on the cell geometry and specific operating conditions. More recently, a general relationship between bubble coverage and the electrochemical cell parameters was proposed, including the bubble shape ($r_R^3/V_r$), current efficiency ($\Phi_B$), gas evolution efficiency ($f_G$), residence time ($t_r$), pressure ($p$), saturation pressure ($p_s$), equivalent radius of detaching bubble ($r_R$), volume at residence time ($V_r$), and stochiometric coefficients of electrons and products ($v_e$, $v_B$).[12] Eq. 15 introduces bubble coverage as function of not only the current density, but also takes into account the effect of other cell parameters which allows to predict bubble coverage under multiple operating conditions:[12]

$$\sigma = \frac{\pi r_R^3}{2V_r} r_R \frac{t_r}{r_R^2} f_G \Phi_B \frac{I/A \; RT}{(v_e/v_B)Fp - p_s} \qquad \text{(Eq. 55)}$$

## 3.2 Bubble Effects on Ohmic Overpotential

The ohmic overpotential arises from the resistance to flow of current between the electrodes. This resistance contains both an ionic component from the flow of charged chemical species through the electrolyte and an electronic component from the flow of electrons through the external circuit. In electrochemical systems, the electronic contribution is often a small fraction of the overall ohmic losses due to the significant difference between the conductivity of electrical conductors and electrolytes.[114] This can be different for systems composed of thin electrodes (<100's nm), such as electrochemical microsystems, where the resistance of the electrode circuits can be comparable to that of the electrolyte.[115, 116, 117, 118, 119, 120, 121] For the purpose of this review, we will focus on processes taking place in the electrolyte, where Ohm's Law can be used to describe (Eq. 16) the ionic current distribution, leading to an integrated expression for the ohmic overpotential,

$$\eta_{ohm} = \int_a^c \frac{i}{\kappa} + \frac{F}{\kappa}\sum_m z_m D_m \nabla c_m \, d\bar{r} \quad \text{(Eq. 16)}$$

where the electrolyte conductivity is $\kappa = F^2 \sum_m z_m^2 u_m c_m$, $z_m$ is the ion valence, $u_m$ is the mobility, $c_m$ is the concentration, $D_m$ is the diffusion coefficient of species $m$, and $\bar{r}$ is the position vector between anode and cathode. In gas evolving systems, bubbles attached to the electrode surface or suspended in the electrolyte will affect the ohmic overpotential differently.[122] Surface-attached bubbles will not only impose an added resistance by preventing current to reach a portion of the electrocatalytic area, but they will induce a non-uniform current density distribution in the area adjacent to the bubbles, affecting the charge transfer and overall overpotential losses in those regions.[103, 105, 108, 122] Figure 7 (right) shows how current density increases around a bubble and how in more tightly-packed configurations of attached bubbles the non-uniformity of the current distribution is more pronounced.[105]

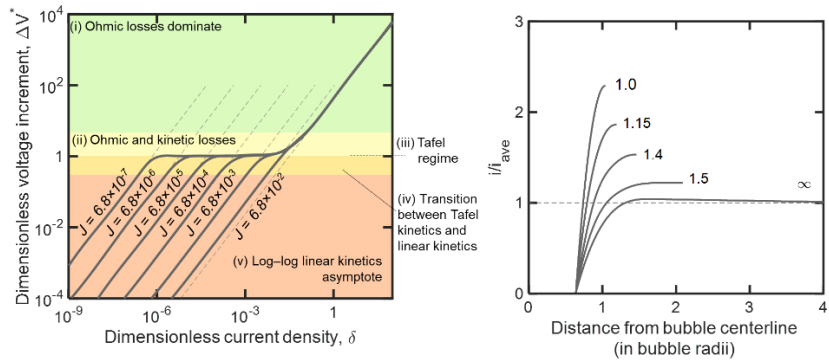

**Figure 7. (left) Dimensionless voltage drop ($\Delta V^*$) due to bubbles attached to electrodes as function of dimensionless current density ($\delta$) for various dimensionless exchange-current densities ($J$).** For moderate exchange current densities, at high $\delta$ the potential losses follow the ohmic resistance asymptote, until reaching a transition to the Tafel regime where activation losses dominate. In the Tafel regime, kinetic losses are only a function of $\sigma$ and $\Delta V^*$ follows a horizontal asymptote. Continuing to lower $\delta$ potential losses transition between Tafel kinetics and log-log linear behavior finally reaching the linear kinetic asymptote. **(right) Modeled current distribution in the area adjacent to a bubble on an electrode surface with a contact angle of 40⁰, for different values of inter-bubble spacing.** Dimensionless quantities are defined as it follows: $\delta = (\alpha_a + \alpha_c)\frac{i_{AVG} r_b}{\frac{RT}{F}\kappa}$, $J = (\alpha_a + \alpha_c)\frac{i_0^o r}{\frac{RT}{F}\kappa}$, $\Delta V^* = \frac{\Delta V}{\frac{RT}{F}}$, where $\alpha_{a/c}$ are the anodic/cathodic, $i_{AVG}$ the average current density at the electrode surface, $i_0^o$ the exchange current density with each reacting species at its bulk concentration, $r_b$ the bubble radius, $\kappa$ the conductivity of the electrolyte, $\Delta V$ the voltage drop due to attached bubbles in the electrolyte, F Faraday's constant, R the universal gas constant and T the temperature of the system.[105] Adapted with permission from Dukovic, J. and C.W. Tobias, *The Influence of Attached Bubbles on Potential Drop and Current Distribution at Gas-Evolving Electrodes.* Journal of The Electrochemical Society, 1987. **134**(2): p. 331-343. Copyright 1987, The Electrochemical Society.

Impedance measurements on a sphere simulating a "frozen" bubble on the electrode surface allowed to determine the increments in charge transfer resistance ($\Delta R_t$), polarization resistance ($\Delta R_p$) and electrolyte resistance ($\Delta R_e$), all of which showed a linear dependence with the square of the diameter of the sphere, demonstrating that ohmic and activation losses increase with the size of bubbles.[108, 123] Bubbles dispersed in the bulk phase also influence the ohmic overpotential by reducing the number of available pathways for ions to migrate and thus lowering the effective conductivity of the electrolyte.[106, 124] Empirical models have been developed describing the relation between the bulk void fraction with the effective conductivity of an electrolyte.[106, 124, 125] Taking Maxwell's and Bruggemann's relations for conductivity as a function of void fraction, an expression that correlates void fraction with relative conductivity of gas dispersions has been proposed.[126, 127] Such expression closely agrees with experimental results and is valid for a range of gas void fraction relevant to water electrolysis, $\epsilon = 0$-$0.12$:[106, 126, 127]

$$\frac{\kappa}{\kappa_0} = (1-\epsilon)^{3/2} \qquad \text{(Eq. 6)}$$

where $\kappa$ and $\kappa_0$ are respectively the conductivity of the electrolyte in the presence and absence of bubbles and $\epsilon$ is the gas void fraction.

The ohmic overpotential dynamically evolves as bubbles nucleate, grow and detach from the electrode surface. These dynamics periodically affect the overall resistance of the electrolyte and thus the ohmic losses in the cell.[123] The implementation of flowing electrolytes has shown to mitigate the potential losses due to dispersed bubbles, while increases in current density enhances such effect due to rapid generation of bubbles.[3, 106, 125, 128] Furthermore, for the case of vertical electrodes where gravity plays a role, the changes in pressure along the length of the electrode results in higher gas volumetric fraction as a function of height, resulting in an uneven local conductivity of the electrolyte and thus a non-uniform current distribution.[110, 123, 125, 129]

### 3.3 Bubble Effects on Concentration Overpotential

Concentration gradients of reactants, intermediates and products in the interelectrode space represent an additional source for potential losses in an electrochemical system.
An expression for the concentration overpotential can be derived from the Nernst equation (Eq. 15) to express the potential losses related to the concentration of reactants and products at both the anode and cathode in an acidic environment.[102]

$$\eta_{conc} = \frac{RT}{2F} \ln\left(\frac{[H^+]_a^2 \, p_{H_2,c} \, p_{O_2,a}^{0.5}}{[H^+]_c^2}\right) \qquad \text{(Eq. 7)}$$

At high current densities the transport of the species involved in the reaction becomes the limiting step of the overall reaction. When an electrochemical reaction takes place in a single aqueous phase, the accumulation of products at the electrocatalytic area will shift the thermodynamic equilibrium towards the reactants (Eq. 15), causing an increase in the concentration overpotential (Figure 8).[102, 130]

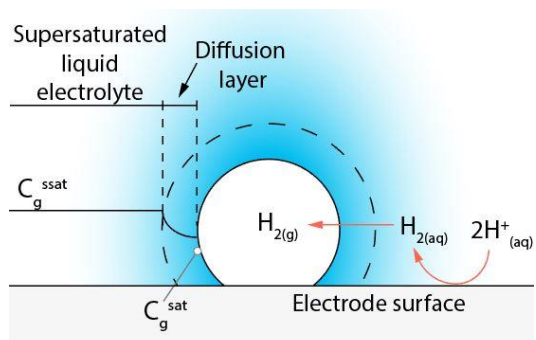

**Figure 8. Concentration profile of gaseous products around a growing bubble.** Dissolved $H_2$ diffuses to the gas phase decreasing supersaturation levels in the electrolyte, modifying concentration gradients and thus affecting the concentration component of the overpotential of an electrochemical cell.

However, when a second phase appears in the form of bubbles on or near the electrodes, they will continue to grow as the dissolved gases diffuse from the electrolyte, reducing the supersaturation of products in the electrolyte which again disturbs the equilibrium towards the products, ultimately reducing the concentration overpotential.[103, 131, 132, 133] A recent study has shown the decoupling the concentration potential from the ohmic and activation by introducing superhydrophobic pits surrounded by a ring shaped microelectrode. Bubbles preferentially nucleated on these hydrophobic sites, preventing the masking of the electrocatalytic area and in consequence removing completely their effect on the kinetic overpotential and significantly the ohmic overpotential.[134] In chronoamperometric experiment, periodic overpotential drops were found to correspond to each bubble evolution event, demonstrating the positive effect of bubbles as they act as gas reservoirs and decrease supersaturation levels near the electrode surface.

Bubbles also affect the concentration component of the overpotential as they grow, causing convective flows that can enhance mass transport limitations, reducing concentration gradients and potentially improving the efficiency of the system.[105, 123] This enhancement is more pronounced when bubbles detach from the electrodes introducing turbulence.[24, 107, 123] Numerical studies on the effect of attached bubbles in hydrogen evolution electrodes showed that concentration overpotential can significantly counteract the activation and resistant components of bubble induced potential losses at moderately slow to fast kinetics, in some cases resulting in a net depolarization of the electrode.[105, 123]

**Table 1. Summary of possible bubble induced effects on each of the overpotential components in an electrochemical gas-evolving system**

| Overpotential component | Typical effect of bubbles |
|---|---|
| Activation | Attached bubbles increase the overpotential due to masking of the electrodes and decrease of the effective electrocatalytic area |
| Ohmic | Attached and free bubbles increase the overpotential due to a blockage of the ion pathways available for current transport. |
| Concentration | Bubbles may decrease the concentration overpotential by absorbing dissolved gas products and decreasing supersaturation levels in the electrolyte<br><br>Bubble detachment and, in some measure, bubble growth, introduce convective flows that disrupt pH gradients which can have a depolarization effect, decreasing the overpotential |

### 3.4 Bubble-induced Energy Losses in Photoelectrodes

Bubbles can induce light-scattering events which affect light absorption in water splitting photoelectrodes. Calculations of the light pathways through bubbles based on Snell's law can be used to estimate the effects on the photocurrent in light absorbers.[135, 136] Experimentally, a few studies have addressed this issue.[135, 136, 137, 138, 139] For example, the bubble effect on illumination has been studied on $TiO_2$ nanorod arrays where oxygen bubbles increased local photocurrents due to concentration of incident light at the three-phase contact line.[136] As the bubble grows, the area surrounding its phase boundary increases and thus the photocurrent enhancement is reduced. A more recent study provides insights into the influence of single bubbles on the photocurrent on planar silicon (Si) photocathodes.[135] Using light tracing models in combination with scanning photocurrent microscopy measurements it was shown that for large bubbles (~ 1000 $\mu$m in diameter) photocurrent losses as high as 23% are observed. Another experimental work reported optical losses of ~5% for an operating current density of 8 mA cm$^{-2}$ in vertical electrodes where a plume of bubbles scattered incoming light.[138] On the other hand, small bubbles (~ 150 $\mu$m) only result in a 2% loss in photocurrent, due to their reduced total internal reflection. Lastly, interactions between multiple bubbles slightly increase the photocurrent losses. Using numerical simulations, it has been demonstrated that the presence of multiple bubbles (sizes ranging from 100 $\mu$m to 800 $\mu$m in diameter) can cause an additional 2% photocurrent loss when compared to the effect of a single bubble.[140]

### 3.5 Bubble-Induced Thermal Losses

Bubbles can also form due to overheating of the electrode surface, further affecting the overpotential contribution. When electrochemical systems are operated at high current densities, thermalized energy losses can lead to temperatures that exceed the boiling point of the electrolyte and in consequence generate bubbles.[141] This thermal evolution of bubbles often occurs near the electrode surface where the increase in electrical resistance caused by evolving electrolytic bubbles leads to boiling; a phenomenon known as the "aqueous anode effect".[141]

### 3.6 Overpotential Fluctuations by Evolving Bubbles

Properties of electrochemical cells (*e.g.*, potential, current density, electrolyte resistance and temperature) dynamically evolve as a result of changes in the active electrode area and dissolved gas concentrations.[108, 123, 142, 143] This dynamic behavior can be exploited to study each of the components of the overpotential. Through electrochemical frequency spectral analysis of the potential and electrolyte resistance, it is possible to isolate the ohmic, activation and concentration components of the total overpotential.[67, 123, 142] This technique provides valuable insights into each of the potential losses at different stages of bubble evolution.[123, 142] Furthermore, power spectral density (PSD) analysis can be used to estimate the size and rates of detaching bubbles by correlating the electrode area covered by a bubble to potential fluctuations by real-time measurements.[67, 134, 144] When operating conditions are such that the system behaves as a "gas oscillator", potential fluctuations occur at regular intervals due the periodic formation, growth and detachment of bubbles. This allows for the study of the bubble dynamics through real time measurements of screening area, contact angle and potential. Figure 9(a) shows the measured overpotential, and below the measured contact angle as function of time. The numbers in the graph correspond to the numbered images in Figure 9(b), depicting the various stages in the bubble evolution. These PSD studies are in good agreement with early numerical estimations on the magnitude of potential losses due to bubbles in electrodes.[103, 105, 108, 123, 142]

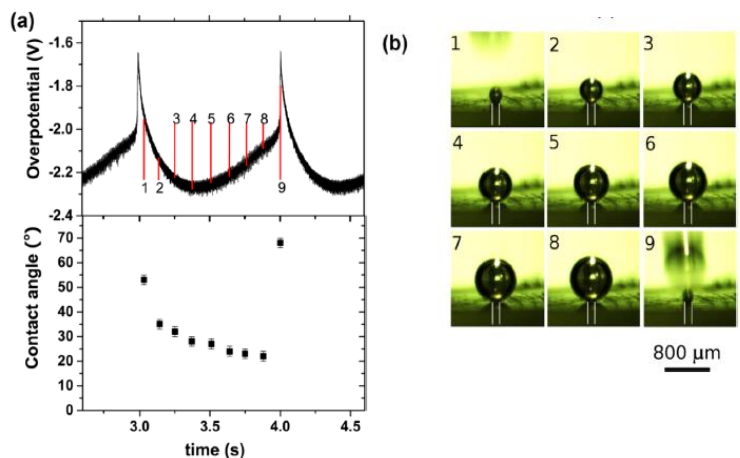

**Figure 9. Bubble evolution on a platinum microelectrode with a diameter of 125 μm.**
(a) The top figure shows the overpotential as function of time during the evolution of a single bubble. The bottom figure shows the contact angle during the bubble evolution as function of time. (b) Images taken during the bubble evolution, where the numbers in the images indicate their time in the overpotential measurement shown in (a). The electrode is indicated by the white lines. Republished with permission from Fernandez, D., et al., *Bubble formation at a gas-evolving microelectrode.* Langmuir, 2014. **30**: p. 13065-13074. Copyright 2014, American Chemical Society.[67]

## 4. Methods to Mitigate or Exploit Bubble Induced Phenomena

Several techniques are available for the removal of bubbles which can be grouped into passive or active. In the former, no additional energy source is needed to promote a change in bubble evolution behavior, whereas in the latter some external source for actuation is employed.

### 4.1 Passive methods

A simple example of spontaneous passive removal are bubble coalescence events, where bubbles are mechanically forced to detach from the surface by their expanding boundaries.[60, 63] Additionally, liquid convection resulting from other detaching bubbles can promote an earlier detachment than theoretically predicted for stagnant electrolytes.[7, 23, 62] The following subsections go beyond these spontaneous removal methods and cover engineering strategies for passive bubble removal.

#### *4.1.1 Geometrical Approaches*

One of the most energy efficient approaches to avoid gas-blocking of active electrocatalytic sites is the implementation of tapered or expanding channel geometries to impose a "self-pumping" effect without external power.[145, 146, 147] This principle is based on capillary pressure gradients that evolve when bubbles are deformed by a tapered channel.[148] Another effective geometrical approach involves the use of hemispherical channels and wells which can capture traveling bubbles in a highly gas permeable microchannel (*e.g.*, Poly-dimethyl siloxane, PDMS, channels), such that the bubbles trapped spontaneously dissipate.[149] This approach is limited by the permeability of the gas in the channel material and is effective when the gas evolution rate is slow.

Passive methods have been implemented in electrochemical systems such as Direct Methanol Fuel Cell (DMFC), demonstrating a >3.5x increase in methanol consumption.[150] Similarly for the case of a hydrogen generator, self-circulation of electrolyte can be implemented to promote directional growth and selective venting of hydrogen bubbles in micro-channels, causing flow mix and fresh solution incoming without extra power required.[151] When coupled with a fuel cell, such mechanism ensures that hydrogen is supplied as demanded by its consumption rate in a fuel cell.

### *4.1.2 Bubble Directed Nucleation Away from Electrocatalytic Sites*

Bubble nucleation sites can be controlled by defining hydrophobic sites (contact angle θ > 90°).[79, 152, 153] Hydrophobic sites are prone to the formation of a gas phase due to the reduced activation energy barrier for bubble nucleation.[47] A recent study explored bubble nucleation and growth on superhydrophobic pits micromachined in silicon electrodes, over which a native $SiO_2$ layer grows.[29, 154] Three distinct regimes were found: first, the bubble growth slows down with each new bubble in the series due to the depletion of the newly-formed concentration boundary layer. Next, the growth rate increases due to a local increase of gas supersaturation due to the continuous gas production, which at last levels off to an approximate steady growth rate. The control on nucleation sites described in these studies demonstrated the possibility to nucleate bubbles away from the electrodes and may be used to minimize bubble-induced energy losses in electrochemical devices.

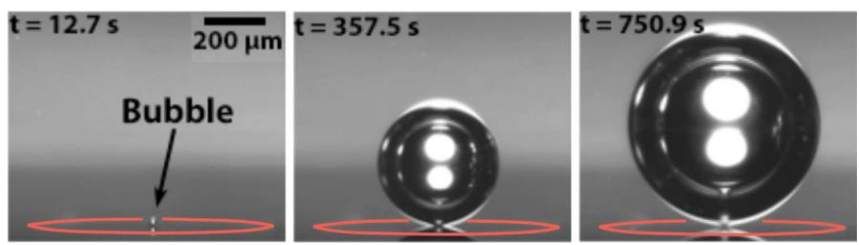

**Figure 10. Bubble evolution at a hydrophobic pit, surrounded by a ring microelectrode.** Sequence of images, taken at a time t since the start of electrolysis, showing a hydrogen bubble nucleating and growing from the hydrophobic pit on a $SiO_2$ substrate. The ring electrode encircling the pit has been highlighted; the current density is 39.3 mA cm$^{-2}$. Reprinted with permission from Peñas, P., et al., *Decoupling Gas Evolution from Water-Splitting Electrodes. Journal of The Electrochemical Society*, 2019. **166**(15): p. H769-H776. Copyright 2019, The Electrochemical Society.

In yet another novel design, a ring microelectrode encircling a hydrophobic microcavity under alkaline water electrolysis conditions, was effective in avoiding bubble coverage.[134] The chronopotentiometric fluctuations of the cell were found to be weaker than in conventional microelectrodes. Numerical transport modeling helped explaining how bubbles forming at the cavity reduce the concentration overpotential by lowering the surrounding concentration of dissolved gas. However, it was found that this architecture can aggravate the ohmic overpotential by blocking ion-conduction pathways.

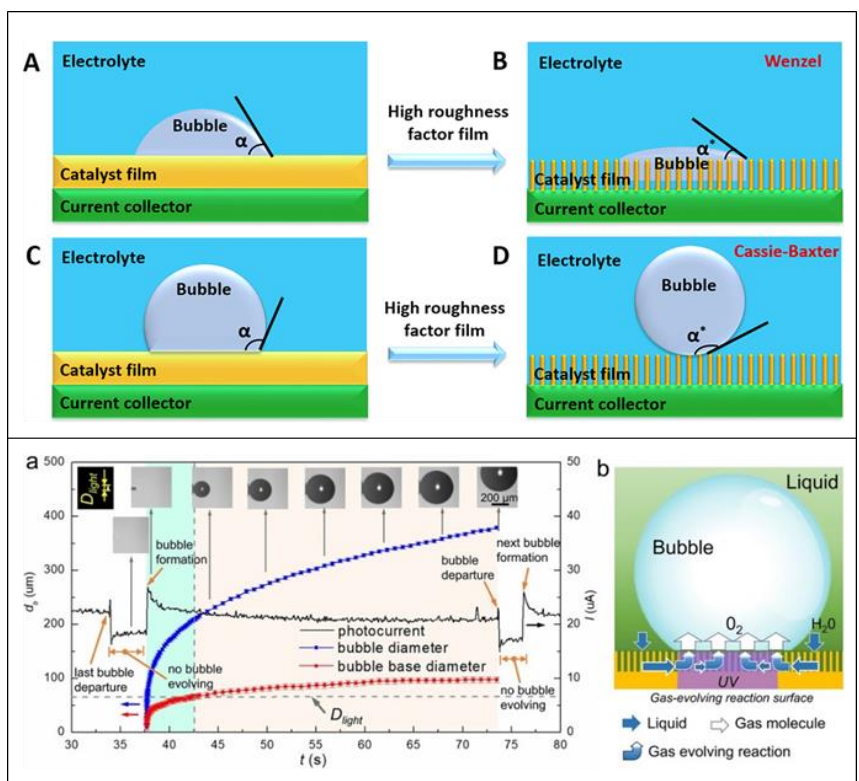

**Figure 11. (top) Schematic illustration of how the surface roughness affecting the bubble contact angle at given intrinsic aerophilicity:** (A) flat aerophilic surface, (B) rough aerophilic surface, (C) flat aerophobic surface, and (D) rough aerophobic surface. (Permission pending) from Xu, W., et al., *Superwetting Electrodes for Gas-Involving Electrocatalysis.* Accounts of Chemical Research, 2018. **51**(7): p. 1590-1598. **(bottom) (a) PEC performance and bubble behaviors during a single bubble cycle (at 10.5 mW light irradiation and no bias potential).** The insets are the side view image of the light spot (leftmost) and the growing bubble. **(b) The liquid film mechanism: The liquid of $H_2O$/$OH^-$ (blue arrows) is imbibed into the bubble base, forming a reactive liquid film.** The liquid film participates in the gas evolving reaction and the generated gas injects directly into the bubble (white arrows), thus sustaining the PEC reaction. (permission pending) . Reprinted with permission from Xu, W., et al., *Superwetting Electrodes for Gas-Involving Electrocatalysis.* Accounts of Chemical Research, 2018. **51**(7): p. 1590-1598. Copyright 2018 American Chemical Society

Structuring the electrode can promote bubble detachment and bursting. The design of superhydrophobic structures that emulated a "self-cleaning" lotus leaf was proven to enhance bubble detachment.[155] In this study, air bubbles were found to burst within a period of 220 ms on microstructured superhydrophobic surfaces, and to detach faster, within 13 ms on similar micro/nanostructured surfaces, demonstrating the importance of the height, width and spacing of hierarchical structures for bubble busting. In another work, nanoarray-based surface engineering technology were implemented to achieve superwetting properties, as roughness of nanoarray architectures, was reported to be a critical factor for constructing superaerophobic and superaerophilic surfaces (Figure 11, top).[156] This study further analyzes the feasibility of superwetting electrodes for enhancing the performances of gas-evolving electrochemical reactions. Moreover, nano-scale capillarity has shown to improve the performance of photoelectrochemical (PEC) water splitting due to a reduction of the blockage of active sites at the bubble base, confirmed by high-speed microscopy.[157] The PEC showed photocurrent of ~20μA in the presence of an attached bubble, which was even higher than that of the free photoelectrode (Figure 11, bottom)

### *4.1.3 Effects of Electrolyte Formulation*

The composition of the electrolyte can affect the properties of bubbles (*e.g.*, surface tension) in electrochemical systems. The detachment radii of bubbles in "surfactant free" electrolytes has been shown to decrease with pH.[65] Addition of Sodium dodecyl sulfate (SDS) or Dodecyltrimethylammonium bromide (DTAB) results in bubbles detaching at smaller radii, due to a decrease in Coulombic forces caused by the adsorption of charged surfactant molecules on the bubble and electrode surfaces.[158] While adding surfactants can help mitigate bubble induced energy losses, it also adds additional complexity due to the possible involvement of these molecules in the electrode reactions. Gaining a better understanding on the effects of electrolyte formulation on electrode processes bears great potential for the improvement of the selectivity and energy efficiency of electrochemical reactors beyond their effects on bubble evolution.

**4.2 Active Methods**

Active methods make use of external forces to prevent bubble nucleation or induce early detachment, mitigating the impact of bubbles in the overpotential of the system. In this section we will describe the most popular strategies, based on the use of flow, magnetic and acoustic fields. Additional methods have been devised in the framework of process intensification to enhance electrochemical processes that involve gaseous products, including centrifugal fields in rotating electrodes.[159, 160]

*4.2.1 Flow Fields*

Shear flow over bubbles can induce early detachment and smaller detachment radii.[85, 161] The bubble detachment from a needle with flow parallel to it has been reported to be smaller than the Fritz radius (see Eq. 9). It has been demonstrated that nucleation rate (frequency of bubble production) and the volume of the detaching bubbles show a non-linear dependence on the shear rate, in particular, higher flow rates over artificial nucleation sites lead to detachment of smaller gas bubbles.[100, 162] This general and intuitive behavior was also observed in a micro-electrolyzer for water splitting.[163] It was also observed that with increased applied voltage, the bubbles detach at smaller radii, attributed to the influence of convection induced by detaching bubbles and electrostatic repulsions between the bubbles and the electrode. More recently, membrane-less architectures for microfluidic electrolyzers enabled by 3D printing, demonstrated the effectiveness of inertial fluidic forces, even in millifluidic regimes.[3, 164, 165]

*4.2.2 Magnetic Fields*

Magnetic fields can induce a Lorentz force on the electrolyte,[166, 167] generating convection that enhances mass transfer, and lowers ohmic and concentration overpotentials. Furthermore, convective flow can favor the desorption of bubbles and reduce their coverage in the electrodes.[168, 169, 170] Reduced overpotentials have been demonstrated for HER electrodes with magnetic fields applied perpendicular to them.[171] A frequency analysis of the noise spectrum during hydrogen evolution demonstrated that a magnetic field of 1.5 T decreased the overpotential for hydrogen formation by 10% - a magnitude comparable to mechanical agitation. Given that magnetic fields can have strong effects on electrolysis efficiency, the magnetization of the electrode material can play a key role in enhancing electrolysis. It has been demonstrated that using ferromagnetic materials can result in larger efficiency improvements than paramagnetic materials, and diamagnetic materials.[167] Furthermore, it was found that a shorter interelectrode distance resulted in more significant magnetic field enhancements.

*4.2.3 Acoustic Fields*

The field of sonoelectrochemistry studies the effects of combined ultrasonic radiation with electrode processes in electrochemical cells.[172] It is known that ultrasound irradiation can lead to acoustic cavitation (bubbles formed due to acoustic pressure fluctuations), acoustic streaming and electrode erosion.[173, 174] Ultrasound has been used to enhance the rate of electrochemical processes by increasing mass transfer in the liquid electrolyte due to the acoustic streaming, and in some cases by actively removing bubbles from electrodes during reactions.[175, 176, 177, 178, 179, 180, 181] More specifically, the complex phenomena associated to sonication can disrupt the diffusion layer, enhance mass transfer of ions through the double layer, and enhance the activity of the electrode surface (see Figure 12).

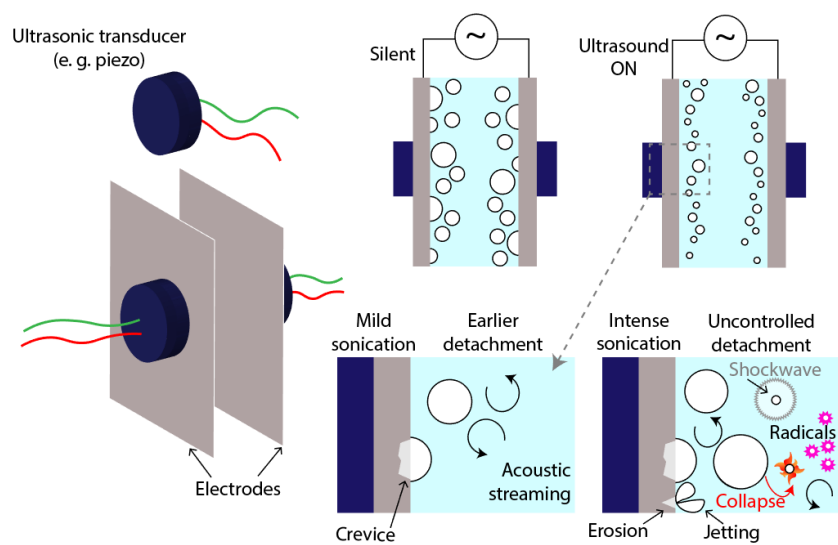

**Figure 12. Schematic representation of a simple geometry where the combination of parallel electrodes and ultrasonic transducers is proposed.** The operation of the electrochemical reactor under silent conditions (without ultrasound) corresponds to the conventional mode. Under ultrasonic irradiation, and below a specific acoustic power (depending on pressure amplitude, frequency, gas concentration in the liquid, etc.) the effects of mild sonication will be positive. However, if cavitation-related phenomena such as jetting and radical production take place, the chaotic nature of bubble collapse will have a detrimental influence on electrochemical processes.

Ultrasound can have effects on electrolysis overpotentials.[176] A reduction in the cell voltage was measured when ultrasound was applied, due to electrode polarization improvements by induced convection as well as a void fraction reduction in the electrolyte. The effect was more pronounced under low electrolyte concentration and high current density conditions, when mass transport limitations are more pronounced. An improvement in hydrogen evolution of 5-18% and energy savings up to 10-25% have been reported. Sonoelectrochemistry has a considerable number of variables that can be adjusted to reach a given goal. Even small changes in the frequency and amplitude of the ultrasound, can affect temperature and gas content in the electrolyte and thus lead to changes in energy conversion efficiency.

## 5. Outlook

Electrochemical systems can play a crucial role in the world's future economy by enabling the electrification of the chemical industry. Throughout this review, we have presented a synthesized account of our current understanding on the impact of bubbles on multiple aspects of electrochemical processes. Despite the substantial literature we have reviewed, we still lack the substantial understanding that would enable us to accurately predict and control the impact of bubbles on electrochemical systems. Paying close attention to electrochemical processes across the scales will position us better to harness the potential of electrochemistry in chemical manufacturing.

We expect that the growing number of experiments and theoretical work focused at understanding bubble-related phenomena at the micro and nanoscale will accelerate the unveiling of new scientific knowledge in this area. For example, faster computers can help simulate detailed physicochemical processes at scales unimaginable a few decades ago. Advanced microfabrication techniques, additive manufacturing and faster cameras, provide a control on geometries and access to phenomena which was impossible to be observe before. These tools can help us gain a better understanding on the interdependent effects that bubbles have on energy conversion efficiencies and electrochemical device performance. Although, this review focused the discussion on water electrolysis as a model reaction, the bubble-induced phenomena described can also inform a multitude of electrochemical systems currently implemented in industry (e.g., chloro-alkali, aluminum production) as well as emerging processes such as $CO_2$ electroreduction or organic electrosynthesis.

**AUTHOR CONTRIBUTIONS**

Conceptualization: D.F.R., H.G., M.M., P.v.d.L., and A.A.; Writing – Original Draft, Review and Editing: P.v.d.L., A.A., M.M. and D.F.R.; Funding Acquisition: D.F.R, H.G. and M.M.; Supervision: D.F.R., H.G. and M.M.

**ACKNOWLEDGMENTS**

We would like to thank our colleagues Pablo Peñas López and Bastian Mei for their help towards our current understanding of this fascinating field.

The collaboration of the author's groups was initiated thanks to the BioSolar Cells program, co-financed by the Dutch Ministry of Economic Affairs (project U1.2).

This work was supported by the Netherlands Centre for Multiscale Catalytic Energy Conversion (MCEC), an NWO Gravitation programme funded by the Ministry of Education, Culture and Science of the government of the Netherlands, and by the Swiss National Science Foundation (SNSF) under the Sinergia grant CRSII5_173860.

**DECLARATION OF INTERESTS**

M.A.M. is a founder and a scientific advisor of Sunthetics, Inc., a start-up company operating in the sustainable chemical manufacturing space.